\documentclass[prb,aps,showpacs,twocolumn]{revtex4}
\usepackage{amssymb}
\usepackage{amsmath}
\usepackage{epsfig}
\usepackage{graphicx}

\begin{document}

\title{First-principles calculations of current-induced spin-transfer torques in magnetic domain walls}

\author{Ling Tang, Zhijun Xu and Zejin Yang}

\affiliation{Department of Applied Physics, Zhejiang University of
Technology, Hangzhou 310023, P.~R.~China}

\begin{abstract}
Current-induced spin-transfer torques (STTs) have been studied in
Fe, Co and Ni domain walls (DWs) by the method based on the
first-principles noncollinear calculations of scattering wave
functions expanded in the tight-binding linearized muffin-tin
orbital (TB-LMTO) basis. The results show that the out-of-plane
component of nonadiabatic STT in Fe DW has localized form, which is
in contrast to the typical nonlocal oscillating nonadiabatic torques
obtained in Co and Ni DWs. Meanwhile, the degree of nonadiabaticity
in STT is also much greater for Fe DW. Further, our results
demonstrate that compared to the well-known first-order nonadiabatic
STT, the torque in the third-order spatial derivative of local spin
can better describe the distribution of localized nonadiabatic STT
in Fe DW. The dynamics of local spin driven by this third-order
torques in Fe DW have been investigated by the
Landau-Lifshitz-Gilbert (LLG) equation. The calculated results show
that with the same amplitude of STTs the DW velocity induced by this
third-order term is about half of the wall speed for the case of the
first-order nonadiabatic STT.
\end{abstract}

\maketitle

\section{Introduction}

The dynamics of the magnetic domain wall (DW) driven by electric
current-induced spin-transfer torque (STT) is under extensive both
experimental\cite{exper1,exper2,exper3,exper4,exper5,exper6,exper7,exper8,exper9}
and
theoretical\cite{XRWang-prl11,Chureemart-prb11,fert-prb13,tatara-pr08,MacDonald-prb09,Gilmore-prb11,Imamura-prb09}
investigations in recent years. As known to all, in general the DW
motion can be driven by an external magnetic field\cite{Walker}
and/or spin-polarized electric current.\cite{Slonczewski,Berger} For
the magnetic field-driven case, although the well-known Walker's
theory\cite{Walker} is usually used to understand the phenomena of
DW movement induced by magnetic field, the origin of this motion is
beyond the scope of Walker's theory. However, X. R. Wang \emph{et
al}.\cite{field-driven1,field-driven2} recently found that the
mechanism of DW propagation by external magnetic field in a nanowire
can attribute to the energy dissipation that is owing to Gilbert
damping. On the other hand for the current-induced case, when an
electron passes through a magnetic DW, it will be scattered by the
noncollinear magnetic structure, which results in the phenomenon of
magnetoresistance (MR).\cite{tatara-IJMB} Meanwhile, the conduction
electron can also transfer the spin angular momentum to the local
spin when it flows through the DW. So a spin-transfer torque will
exert on the local magnetic moment and then the electric current can
be used to manipulate the magnetic structure of DW. This phenomenon
of STT including current-induced DW motion was first predicted and
observed in the experiments by L. Berger and
coworkers.\cite{berger1,berger2,berger3} Moreover, this exotic
phenomenon will also have some potential applications such as
magnetic random-access memories (MRAMS),\cite{MRAMS1,MRAMS2}
racetrack memories,\cite{racetrack1,racetrack2} spin transfer nano
oscillators (STNOs)\cite{Berger,STNO2,STNO3} etc., which have
attracted a great deal of attention recently.

For a DW in the adiabatic approximation, the spin of the incident
electrons can be consistently aligned with the local spin moments.
The spatial derivative of this adiabatic spin current yields the
adiabatic STT which is written as $-b_{J}\partial_{y}\mathbf{S}$,
where $\mathbf{S}$ is the local spin of DW and $y$ is coordinate in
transport direction.\cite{SCZhang-prb98,Li-prb04,Li-prl04} Here
$b_{J}$ is parameter with unit [m/s] in proportion to current
density. However, there are always some conduction electron spins
that cannot follow the local spin moments as the width of DW
decreasing. The process that the spin of conduction electron relax
toward the local spin moments will cause a nonadiabatic
torque\cite{Zhang-prl04,Waintal-eul04,Barnes-prl05} called the
$\beta$ term, which is in the form of $-\beta
b_{J}\mathbf{n}\times\partial_{y}\mathbf{S}$, where $\beta$ is the
dimensionless parameter and $\mathbf{n}=\mathbf{S}/|\mathbf{S}|$ is
the unit vector of local spin moments. Further, S. Zhang and Z.
Li\cite{Zhang-prl04} have pointed out that the adiabatic STT only
contributes the initial velocity of DW movement while the
nonadiabatic term mentioned above determines the terminal velocity
observed in experiments.

However, for the relative narrow DW a nonlocal oscillating torques
have been predicted theoretically by several
groups\cite{Waintal-eul04,MD-prb06,Brataas-prl07} and its quantum
origin is similar to the RKKY oscillation.\cite{tatara-pr08}
Moreover, in general the ratio between the maximum of nonadiabatic
and adiabatic STT, such as the coefficient $\beta$ mentioned above,
represents the degree of nonadiabaticity which can determine the
velocity of DW movement.\cite{Zhang-prl04} So one of the purposes in
this paper is to find out whether there is a nonlocal oscillating or
localized torques in real ferromagnetic materials by calculating the
STT of DW using the first-principles method. And the other purpose
is to study the difference in magnitude of nonadiabatic STT among
the traditional ferromagnetic DWs (Fe, Co and Ni).

On the other hand, M. Thorwart and R. Egger\cite{Thorwart-prb07}
have derived a torque in form of the second-order spatial derivative
of the local spin by a gradient expansion scheme. Hence, it is
tempted to introduce a higher-order of nonadiabatic STT such as
$\mathbf{n}\times\partial^{3}_{y}\mathbf{S}$ into the
current-induced DW dynamics.\cite{Wessely-jpc09} So in this paper we
will figure out whether there is higher-order torque in DW for real
ferromagnetic materials. In addition, the effect of such
higher-order STT on the DW movement is also needed for better
understanding of current-induced DW dynamics.

In this paper, we will study the current-induced STT of defect-free
DW in ballistic limit by the first-principles electronic structure
calculations.\cite{shuai-prb08,shuai-prb10,xu-prl08} Our results
show that the nonadiabatic STT of Fe DW has localized form while the
out-of-plane STTs of Co and Ni are typical nonlocal oscillating
torques. The degree of nonadiabaticity is also much greater for Fe
DW and increases exponentially with decreasing the width of DW. In
addition, the results also show that the distribution of our
calculated nonadiabatic STT for Fe DW can well describe by the
third-order spatial derivative term as
$\mathbf{n}\times\partial^{3}_{y}\mathbf{S}$. Finally, the dynamics
of local spin in DW pushed by this third-order STT is simulated
using Landau-Lifshitz-Gilbert (LLG) equation. The obtained time
evolution of DW movement demonstrates that with the same amplitude
of the nonadiabatic STT the velocity owing to the third-order STT
will be reduced to about half of the wall speed induced by the
first-order term.

\section{Method}

\begin{figure}
  \includegraphics[width=8.6cm, bb=12 4 593 181]{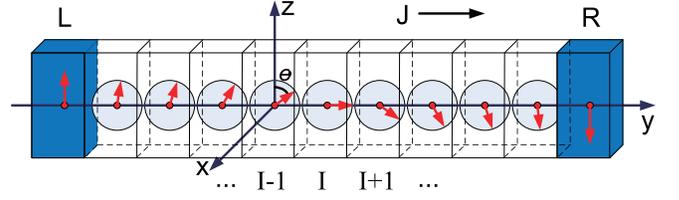}\\
  \caption{Sketch of the completed
scattering problem of layered system for N\'{e}el-type domain wall.
The red arrows distributed in $y$-$z$ plane denote the local spin
moments whose polar angle is $\theta$. Note that the current ($J$)
flows from the left (L) lead to the right (R) lead and $I$ is layer
index. }\label{f1}
\end{figure}

Our calculation of STT for DW is based on the scattering wave
function matching (WFM) method with tight-binding linear muffine-tin
orbital (TB-LMTO) basis.\cite{shuai-prb08,xia-prb06} For a typical
N\'{e}el or Bloch DW, the left and right domains act as leads and
the wall structure is regarded as the scattering region, which
defines the completed scattering problem of layered system as shown
in Fig. \ref{f1}. For this layered system, we assumed that the
magnetization of DW has lattice translation invariant in the plane
perpendicular to transport direction, so the operator and scattering
states can be characterized by a lateral $\mathbf{k}_{\parallel}$
wave vector in two-dimension (2D) Brillouin zone (BZ).

In this paper, the STT is calculated from the spin current, which is
defined as
\begin{equation}
\hat{\mathcal{J}}\equiv \frac{1}{2}\left[
\mathbf{\hat{\sigma}}\otimes
\mathbf{\hat{V}}+\mathbf{\hat{V}}\otimes
\mathbf{\hat{\sigma}}\right] . \label{js-define}
\end{equation}
where $\hat{\sigma}$ is Pauli spin matrix and $\mathbf{\hat{V}}$ is
velocity operator. In the mixed representation for a special
$\mathbf{k}_{\parallel}$ with real space quasi one-dimension (1D)
tight-binding model, the spin current operator from \textbf{R}'th to
\textbf{R}th site is\cite{shuai-prb08}
\begin{equation}
\hat{\mathcal{J}}_{\mathbf{R}^{\prime },\mathbf{R}}\left(
\mathbf{k}_{\parallel }\right) =\underset{LL^{\prime }}{\sum }
\frac{1}{2i\hbar}\left[
\mathbf{\hat{\sigma}\hat{H}}_{\mathbf{R}L,\mathbf{R}^{\prime
}L^{\prime }}^{\mathbf{k}_{\parallel }} +
\mathbf{\hat{H}}_{\mathbf{R}L,\mathbf{R} ^{\prime }L^{\prime
}}^{\mathbf{k}_{\parallel }}\mathbf{\hat{\sigma}} -h.c. \right] .
\label{js-opt}
\end{equation}
where
$\mathbf{\hat{H}}^{\mathbf{k}_{\parallel}}_{\mathbf{R}L,\mathbf{R}L'}$
is noncollinear TB-LMTO Hamiltonian matrix element in the global
quantum axis representation. Here $L\equiv(l,m)$, where $l$ and $m$
are the azimuthal and magnetic quantum numbers respectively.

In order to obtain
$\mathbf{\hat{H}}^{\mathbf{k}_{\parallel}}_{\mathbf{R}L,\mathbf{R}L'}$
of DW, firstly the collinear Hamiltonian from self-consistent
one-electron effective potential of the magnetic materials is
calculated by collinear electronic structure calculation in atom
sphere approximation (ASA).\cite{asa1,Andersen-prl84,Andersen-prb86}
Next, by introducing the rigid potential
approximation,\cite{shuai-prb08,shuai-prb10,tang-mplb08} we rotate
the above Hamiltonian, which is diagonal in $2\times2$ spin space at
local quantum axis representation, by unity rotation matrix
$\hat{U}(\theta,\phi)$ in spin space to construct the noncollinear
Hamiltonian in the global quantum axis representation. Here the
rotation angle $\theta$ is determined by the local spin moments
configuration of DW, which is written as
\begin{equation}
    \theta(y_{\mathbf{R}})=\frac{\pi}{2}+\arcsin[\tanh(\frac{y_{\mathbf{R}}}{\lambda_{\mathrm{DW}}})]\label{lambda}
\end{equation}
where $\lambda_{\mathrm{DW}}$ is the the characteristic length for
DW and $\theta(y_{\mathbf{R}})$ is the polar angle of local spin on
the \textbf{R}th site with position $y_{\mathrm{\mathbf{R}}}$ along
transport direction. Here, the injected current is along fcc(111)
direction for Co and Ni DW and bcc(001) direction for Fe DW. In our
calculation, we choose the azimuthal angle
$\phi(y_{\mathbf{R}})=\pi/2$ for modeling N\'{e}el-type DW as shown
in Fig. \ref{f1}.

Using the WFM method, we can obtain the scattering wave function
corresponding to the noncollinear Hamiltonian of DW. Therefore, for
scattering state with lateral wave vector $\mathbf{k}_{\parallel}$,
the expectation value of STT acting on local spin can be determined
by the difference between the incoming and outgoing spin current on
\textbf{R}th site, i.e.,\cite{shuai-prb08,shuai-prb10}
\begin{equation}
\left\langle \mathbf{\hat{T}}^{s}_{\mathbf{R}}\left(
\mathbf{k}_{\parallel
}\right) \right\rangle  \\
=\underset{\mathbf{R}^{\prime }\in I-1,I}{\sum }\left\langle
\hat{\mathcal{ J}}^{s}_{\mathbf{R}^{\prime },\mathbf{R}}\left(
\mathbf{k}_{\parallel }\right) \right\rangle
-\underset{\mathbf{R}^{\prime }\in I,I+1}{\sum } \left\langle
\hat{\mathcal{J}}^{s}_{\mathbf{R},\mathbf{R}^{\prime }}\left(
\mathbf{k}_{\parallel }\right) \right\rangle . \label{expectation}
\end{equation}
where $I$ is the index of principal layer in quasi-1D model and the
\textbf{R}th site belongs to the layer $I$. Here the superscript
$s=\uparrow,\downarrow$ denotes that the scattering state for
evaluating expectation value is induced by the injected electron in
spin $s$ with respect to the local quantum axis of lead. Then in the
linear response regime, the total torque under a small bias $V_{b}$
is calculated by summing all the STT of $\mathbf{k}_{\parallel}$
states in 2D BZ, which can be written
as\cite{shuai-prb08,shuai-prb10}
\begin{equation}
\mathbf{T}_{\mathbf{R}}(V_{b})=\left(  \frac{\hbar}{2}\right)
\frac{e}{2h}\frac
{V_{b}}{N_{\parallel}}\underset{s,\mathbf{k}_{\parallel}}{\sum}\left[
\left\langle \mathbf{\hat{T}}_{\mathbf{R}}^{s}\left( \mathbf{k}
_{\parallel}\right)  \right\rangle_{\mathcal {L}} -\left\langle
\mathbf{\hat{T}}_{\mathbf{R} }^{s}\left(
\mathbf{k}_{\parallel}\right)  \right\rangle_{\mathcal {R}} \right]
,\label{t-v}%
\end{equation}
where $N_{\parallel}$ is the number of $\mathbf{k}_{\parallel}$
states in 2D BZ at Fermi level. Here $\mathcal {L}$ and $\mathcal
{R}$ denote the STT induced by the left and right incoming electrons
from the lead region respectively. In our calculations, the STTs of
DW are performed with a $480\times480$ $\mathbf{k}_{\|}$-mesh points
in the lateral 2D BZ, which can insure convergence of the results.

\section{Results and Discussion}

\begin{figure}
  \includegraphics[width=8.6cm, bb=14 73 442 318]{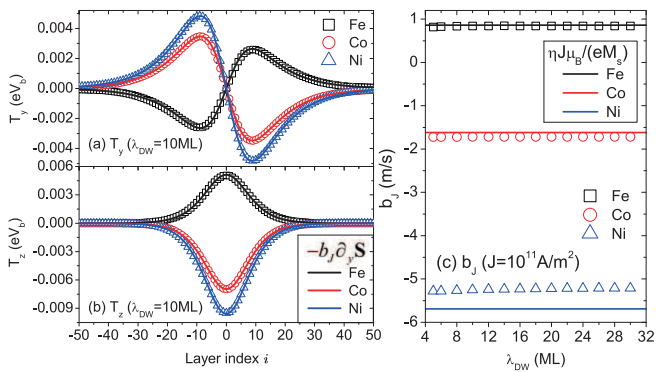}\\
  \caption{The in-plane STT for Fe, Co
and Ni DWs with $\lambda_{\mathrm{DW}}=10$[ML] at unit bias. (a) $y$
component of STT. (b) $z$ component of STT. The solid line denote
the fitting adiabatic STT in form of $-b_{J}\partial_{y}\mathbf{S}$.
It can be seen that our calculated in-plane STTs agree well with the
term of $-b_{J}\partial_{y}\mathbf{S}$. (c) The fitting results of
$b_{J}$ for different DW width at current density
$J=10^{11}[\mathrm{A}/\mathrm{m}^{2}]$. Here the solid line denote
the predicted values of $b_{J}=\eta J\mu_{B}/(eM_{s})$. }\label{f2}
\end{figure}

\begin{table}
\caption{The parameter $b_{J}$ at current density $J=10^{11}$
[A/m$^{2}$] for $\lambda_{\mathrm{DW}}=10$[ML], where $\eta$ is the
calculated polarization
$\eta\equiv(G_{\uparrow}-G_{\downarrow})/(G_{\uparrow}+G_{\downarrow})$,
where $G_{\uparrow(\downarrow)}$ is sharvin conductance of leads for
spin up (down) channel at Fermi level in our transport calculation.
One can see that the values of our fitting $b_{J}$ are close to
$\eta J\mu_{B}/(eM_{s})$.}

\begin{tabular}{ccccccc}
\hline \hline
 DW & $G_{\uparrow}$[$\frac{\mathrm{S}}{\mathrm{m}^{2}}$] & $G_{\downarrow}$[$\frac{\mathrm{S}}{\mathrm{m}^{2}}$] & $\eta$ &   $M_{s}$ [$\frac{A}{m}$] &
 $\frac{\eta J\mu_{B}}{eM_{s}}$ [$\frac{m}{s}$] & $b_{J}$ [$\frac{m}{s}$] \\
\hline
 Fe & $7.78\times10^{14}$ & $4.6\times10^{14}$  & 0.257   & $17.18\times10^{5}$   & 0.866 & 0.84\\
 Co  & $4.59\times10^{14}$ & $10.8\times10^{14}$ & -0.40   & $14.46\times10^{5}$    & -1.60 & -1.72\\
 Ni  & $4.69\times10^{14}$ & $13.4\times10^{14}$ &  -0.48    &  $4.9\times10^{5}$    &  -5.67 & -5.25\\
 \hline
 \hline
\end{tabular}

\end{table}

In this paper, we will investigate the nonadiabatic STT and
corresponding dynamics of DW movement for the typical ferromagnetic
materials Fe, Co and Ni. Firstly, Fig. \ref{f2}(a) and (b) show the
$y$ and $z$ component of STT for DW width
$\lambda_{\mathrm{DW}}=10$[ML]. According to the configuration of
N\'{e}el DW in Fig. \ref{f1}, here the $y$ and $z$ component of
torque is so-called in-plane STT and the $x$ component is the
out-of-plane STT. It can be seen that all these in-plane torques
including Fe, Co and Ni agree well with the prediction of adiabatic
approximation,\cite{SCZhang-prb98,Li-prb04,Li-prl04} which has the
form of $-b_{J}\partial_{y}\mathbf{S}$. It is noted that the STT
shown in Fig. \ref{f2}(a) and (b) is at unit bias and the
coefficient $b_{J}$ is proportional to the current density. So
combined with the corresponding ballistic conductance
$G$[$\mathrm{S}/\mathrm{m}^{2}$] of DW, we can use formula
$-b_{J}\partial_{y}\mathbf{S}$ to fit the in-plane STT and obtain
the coefficient $b_{J}$ at unit current density. Here Table. 1 shows
the fitting results of parameter $b_{J}$ at current density
$J=10^{11}[\mathrm{A}/\mathrm{m}^{2}]$ for
$\lambda_{\mathrm{DW}}=10$[ML]. One can observe that our fitting
coefficients $b_{J}$ for Fe, Co and Ni DWs are close to the values
of $\eta J\mu_{B}/(eM_{s})$, which is the prediction of the
semiclassical transport theory,\cite{Li-prb04,Li-prl04} where $\eta$
is the spin polarization of current, $J$ is current density and
$M_{s}$ is saturated magnetization of ferromagnet. Here note that we
use the calculated polarization
$\eta\equiv(G_{\uparrow}-G_{\downarrow})/(G_{\uparrow}+G_{\downarrow})$,
where $G_{\uparrow(\downarrow)}$ is sharvin conductance of leads for
spin up (down) channel at Fermi level in our transport calculation,
instead of the values extracted from the experiments. The fitting
results of $b_{J}$ for different DW width are also shown in Fig.
\ref{f2}(c). It can be seen that the $b_{J}$ is almost independent
on DW width even for the narrow wall with
$\lambda_{\mathrm{DW}}=5$[ML].

\begin{figure}
  \includegraphics[width=8.6cm, bb=6 80 482 340]{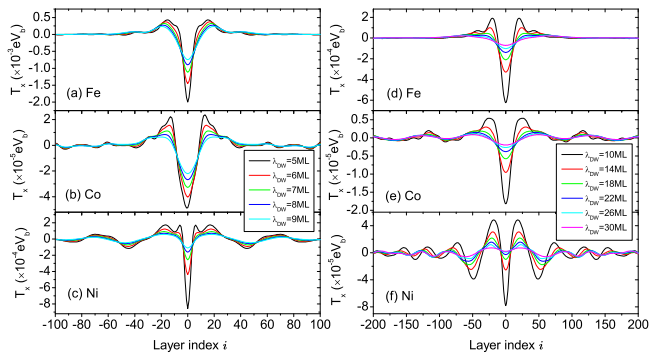}\\
  \caption{The out-of-plane component of
STTs for Fe, Co and Ni DWs with different $\lambda_{\mathrm{DW}}$.
One can observe that the out-of-plane STT of Fe DW has localized
form while the spatial distribution of STTs for Co and Ni DW is
nonlocal oscillating. }\label{f3}
\end{figure}

Next, we will concentrate on the obtained out-of-plane nonadiabatic
STT which is beyond the above in-plane adiabatic STT. Fig. \ref{f3}
shows the spatial distribution of out-of-plane STT for different
width $\lambda_{\mathrm{DW}}$ in Fe, Co and Ni DWs. One can observe
that for Co and Ni DWs in the region far away from the DW center,
where the magnetization is uniform, there is nonlocal oscillation in
the distribution of out-of-plane torque. This nonlocal oscillating
torques have already been discovered by several
groups\cite{Waintal-eul04,MD-prb06,Brataas-prl07} and G. Tatara
\emph{et al}.\cite{tatara-pr08} pointed out that the oscillating
torques can be summed up as a collective force on the DW. In our
calculations, this oscillating out-of-plane STT is contributed by
the propagating states which precess around the local spin moments
and has the form of $e^{i(k^{\uparrow}-k^{\downarrow})y}$, where
$k^{\uparrow(\downarrow)}$ is wave vector in transport direction of
propagating state.\cite{Stiles-prb02} Considering that the summation
of propagating states over the 2D BZ in the transport calculation,
the cancellation effect\cite{Stiles-prb02} in the different
$\mathbf{k}_{\parallel}$ make the oscillating behavior of
out-of-plane torque depend on the shape of Fermi surface. Therefore,
as shown in Fig. \ref{f3} the amplitude and diffusion length of
torque oscillation is significant larger for Ni DW than that for Co
DW, which is similar to the case of layered spin valve
system.\cite{shuai-prb08} For the same reason as above, in Fig.
\ref{f3}(a) and (d) one can see that our calculated out-of-plane STT
of Fe DW has localized form without oscillation far away from DW
center except a small oscillation near the center region.

\begin{figure}
  \includegraphics[width=8.6cm, bb=22 14 448 340]{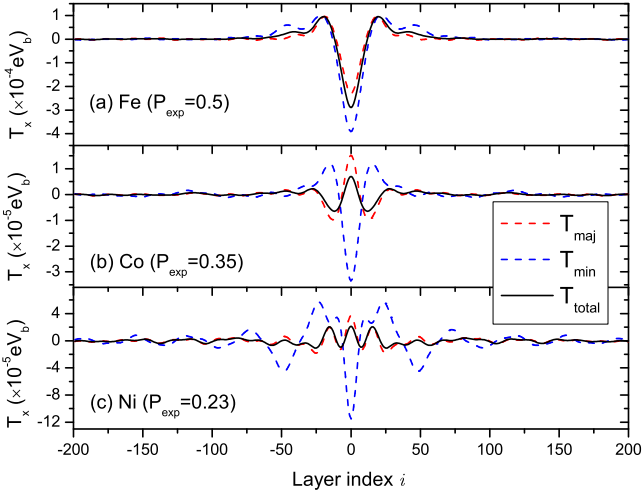}\\
  \caption{The calculated distribution of
out-of-plane STTs for DWs with $\lambda_{\mathrm{DW}}=10$[ML] in the
experimental current spin polarization. The black solid line denotes
the total weighted STT, while the red and blue dash lines denote the
STT induced only by incoming majority and minority electron
respectively. }\label{fig-suppl}
\end{figure}

In our ballistic calculations, the spin polarization of the current
is different with the experimental one. So in order to simulate the
case of STT in experiments, firstly using the Eq. \ref{t-v} we can
obtain the STT induced only by the incoming majority
($\mathbf{T}_{\uparrow}$) and minority ($\mathbf{T}_{\downarrow}$)
electron respectively. Then the total STT can be calculated by
$\mathrm{T}_{\mathrm{total}}=W_{\uparrow}\mathbf{T}_{\uparrow}+W_{\downarrow}\mathbf{T}_{\downarrow}$,
where the weighting factors $W_{\uparrow(\downarrow)}$ is determined
by
$(W_{\uparrow}G_{\uparrow}-W_{\downarrow}G_{\downarrow})/(W_{\uparrow}G_{\uparrow}+W_{\downarrow}G_{\downarrow})=P_{\mathrm{exp}}$.
Here $P_{\mathrm{exp}}$ is the experimental current spin
polarization. Fig. \ref{fig-suppl} shows the distribution of the
out-of-plane STTs for DWs with the experimental current spin
polarization. One can observe that the results do not qualitatively
change at all compared to the ballistic calculations, namely the
out-of-plane STT of Co and Ni DWs have nonlocal oscillating forms
while that of Fe DW has the localized form.

\begin{figure}
  \includegraphics[width=8.6cm, bb=35 9 427 305]{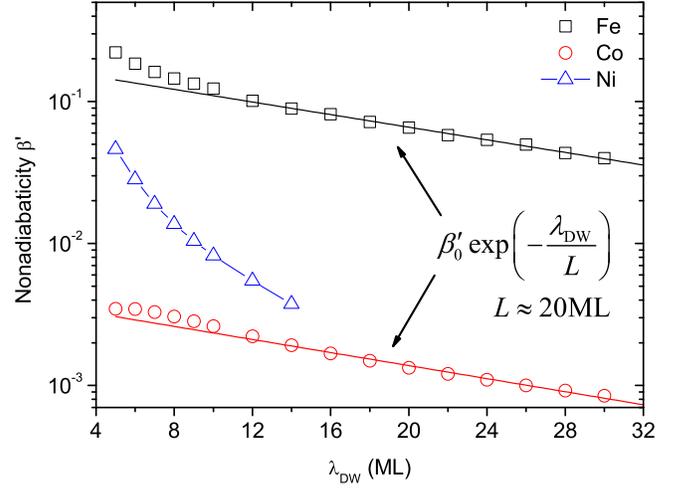}\\
  \caption{The width dependence of
nonadiabaticity for Fe, Co and Ni DWs. The nonadiabaticity $\beta'$
is defined as the ratio between the out-of-plane and in-plane STT at
DW center, namely
$\beta'\equiv|T_{x}/\sqrt{T_{y}^{2}+T_{z}^{2}}|_{y=0}$. It is noted
that for Ni DW with $\lambda_{\mathrm{DW}}>14$[ML] the oscillation
of out-of-plane STT is so larger that the maximum is no longer at
the center of DW. For $\lambda_{\mathrm{DW}}>10$[ML] the
nonadiabaticity of Fe and Co can be well fitted by
$\beta'=\beta_{0}'\exp(-\lambda_{\mathrm{DW}}/L)$, where
$L\approx20$[ML]. }\label{f4}
\end{figure}

For the out-of-plane torque in the form of $-\beta
b_{J}\mathbf{n}\times\partial_{y}\mathbf{S}$, it is believed that
the coefficient $\beta$, which is just the ratio between the maximum
of out-of-plane and in-plane STT, determine the final velocity $v$
of DW motion by the expression\cite{Zhang-prl04} of
$v=b_{J}\beta/\alpha$. And coefficient $\beta$ is usually viewed as
the degree of nonadiabaticity in STT. However, the form of
out-of-plane nonadiabatic STT in our results is not exactly the same
with the $\beta$ term, but we still calculated the ratio between the
out-of-plane and in-plane STT at the center of DW to demonstrate the
degree of nonadiabaticity for DW with different materials. As shown
in Fig. \ref{f4}, one can see that with decreasing the DW width the
nonadiabaticity will increase exponentially, which agree well with
the calculation of free-electron Stoner model.\cite{MD-prb06}
Further, the nonadiabaticity for Co and Fe DWs have the same growth
rate with decreasing $\lambda_{\mathrm{DW}}$. Moreover, one can also
observe that the nonadiabaticity of Fe DW is about one order larger
than that of Co and Ni DWs. In the meantime, the value of
nonadiabaticity for Fe DW even can be reach at a very high value
($\sim0.2$) for the narrow width $\lambda_{\mathrm{DW}}=5$[ML]. In
general, there are two contributions to the nonadiabatic STT in DW.
One is the spin relaxation process of the conducting electron spin
toward local spin, the other is the mistracking between the incoming
conducting electron spin and local spin without any relaxation. In
our calculations, it is noted that there is no any dynamical
relaxation process taken into account, thus only the mistracking
contribution exists, which is similar to the case of free-electron
Stoner model.\cite{MD-prb06} From Fig. \ref{f4}, one can see that
this mistracking contributed STTs for Co and Ni are too small to be
measured in experiments with reasonable DW width. Therefore, it
implied that the relaxation contributed nonadiabatic STT dominates
in Co and Ni DWs.

\begin{figure}
  \includegraphics[width=8.6cm, bb=26 14 448 340]{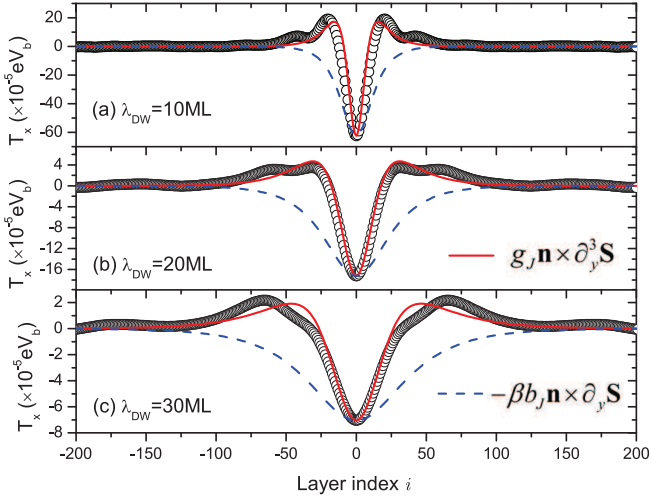}\\
  \caption{The calculated distribution of
out-of-plane STT and the spatial derivative of the local spin for Fe
DW. Here the hollow circles are obtained STTs from the
first-principles calculations. The blue dash line and red solid line
denote the fitting results of $\beta$ term and the torques in the
third-order of the spatial derivative respectively. It shows that
the term of $g_{J}\mathbf{n}\times\partial^{3}_{y}\mathbf{S}$ can
well describe the calculated out-of-plane STT of Fe DW. }\label{f5}
\end{figure}

Fig. \ref{f5} shows the comparison between the calculated
out-of-plane STT and the torque of spatial derivative of the local
spin. Here the torques in first- and third-order derivative are both
depicted and their maximums are fitted to the calculated STT at
center. At first glance, one can see that the configuration of the
$\beta$ term is much different with the calculated results from Fig.
\ref{f5}. In particular, the width of STT peak for $\beta$ term at
the DW center is much larger than that of the calculated STT. Hence,
it suggested that this deviation may have a significant effect on
the velocity of DW movement. However, the study of the DW movement
using the first-principles calculated STT directly will cost too
much computing time so that we can hardly obtain the results for a
long time intervals. So in order to investigate the time evolution
of DW more efficiently, here we will model a simple analytical
nonadiabatic STT term substituted into the equations of DW dynamical
evolution. As shown in Fig. \ref{f5}, we found that compared to the
$\beta$ term, the nonadiabatic STT in the form of
$g_{J}\mathbf{n}\times\partial^{3}_{y}\mathbf{S}$, which is
proportional to the third-order spatial derivative of the local
spin, can well describe the obtained out-of-plane STT in Fe DW. Here
$g_{J}$ is the parameter in proportion to current density. The
origin of this nonadiabatic STT may be similar with the $\beta$
term. According to the gradient expansion scheme for
STT\cite{Thorwart-prb07}, the first-order expansion will result in
both adiabatic STT and nonadiabatic $\beta$ term. Meanwhile, the
higher-order expansion will lead to the second-order
STT.\cite{Thorwart-prb07} Therefore, based on the gradient expansion
scheme, it is reasonable to obtain the torques in the third-order
spatial derivative in our calculations.

On the other hand, at two sides of Fe DW center region there is
still small difference between the calculated and the third-order
STT, where the calculated STT also has small oscillating form. It
implied that the reason of our calculated STT in Fe DW being well
fitted by the third-order term maybe is due to the dephasing effect,
where the oscillating STT can be canceled each other by summing the
$\mathbf{k}_{\parallel}$ states at Fermi surface. For Co and Ni DW
the majority Fermi surface is similar to that of free-electron thus
the dephasing effect is very weak, which makes the nonadiabatic STT
still has oscillating form\cite{MD-prb06} as shown in Fig.
\ref{fig-suppl}(b) and (c). However, the Fermi surface of Fe is far
beyond the free-electron-like one so the dephasing effect is much
stronger in Fe DW. Meanwhile, the dephasing effect is not effective
near the DW center and increases its effectiveness relatively
rapidly with increasing distance from the DW center. Therefore, the
corresponding nonadiabatic STT has the form of the third-order term.
In particular, when the distance from DW center is not very large,
the dephasing has not taken fully effect in such regions, so the
oscillating behavior of STT will still emerge as shown in Fig.
\ref{f5}. Nevertheless, due to the dynamics of DW is mainly
attribute to the STT in DW center, which is well fitted by the
third-order STT as shown in Fig. \ref{f5}, in this paper we will
focus on how the nonadiabatic STT term in the form of
$g_{J}\mathbf{n}\times\partial^{3}_{y}\mathbf{S}$ will influence the
velocity of DW movement.

\begin{figure}
  \includegraphics[width=8.6cm, bb=1 9 468 344]{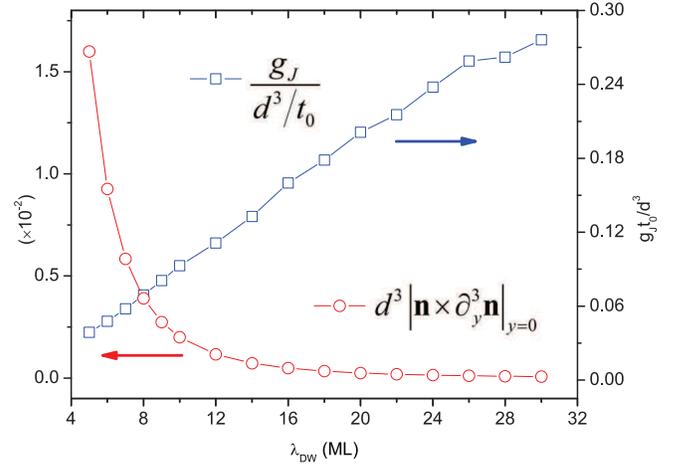}\\
  \caption{The fitting results of
dimensionless coefficient $g_{J}t_{0}/d^{3}$ for Fe DW with
different width at current density
$J=10^{11}[\mathrm{A}/\mathrm{m}^{2}]$, where $d$ is the distance
between the monolayers and the unit time is $t_{0}=(4\pi
M_{s}\gamma)^{-1}$. The dependence of the third-order of spatial
derivative at DW center ($y=0$) on width $\lambda_{\mathrm{DW}}$ is
also depicted in this figure. It can be seen that although $g_{J}$
grows linearly with increasing $\lambda_{\mathrm{DW}}$, the
amplitude of the third-order torque still decays due to the term of
$\mathbf{n}\times\partial^{3}_{y}\mathbf{n}$. }\label{f6}
\end{figure}

For Fe DW with different width $\lambda_{\mathrm{DW}}$ at current
density $J=10^{11}[\mathrm{A}/\mathrm{m}^{2}]$, the fitting results
of dimensionless coefficient $g_{J}t_{0}/d^{3}$ is shown in Fig.
\ref{f6}, where $g_{J}$ is in unit of [m$^{3}$/s], $d=1.43${\AA} is
the distance between the monolayers and the unit time is
$t_{0}=(4\pi M_{s}\gamma)^{-1}=2.63[\mathrm{ps}]$. Here, our
obtained dimensionless coefficient $g_{J}t_{0}/d^{3}$ can provide
the magnitude of the third-order torque which maybe included in some
micrmagnetic simulations such as OOMMF. In addition, it is noted
that the factor of the third-order torque $g_{J}$ increases linearly
with increasing $\lambda_{\mathrm{DW}}$. However, the exponential
decay of $\mathbf{n}\times\partial^{3}_{y}\mathbf{n}$ term will lead
to the obtained third-order torque still decreasing with increasing
$\lambda_{\mathrm{DW}}$.

\begin{figure}
  \includegraphics[width=8.6cm, bb=27 14 434 308]{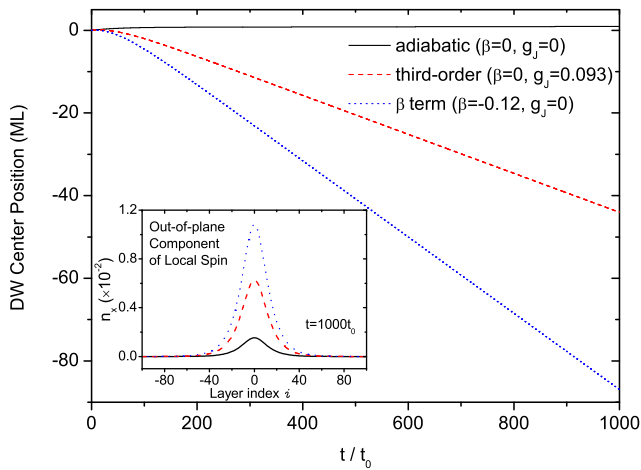}\\
  \caption{The time evolution of Fe DW
center with $\lambda_{\mathrm{DW}}=10$[ML] at
$J=10^{11}[\mathrm{A}/\mathrm{m}^{2}]$ below the critical current
density. Here the unit time is $t_{0}=(4\pi M_{s}\gamma)^{-1}$. One
can observe that the velocity driven by the third-order nonadiabatic
STT is about half of the wall speed driven by the first-order
$\beta$ term. The inset shows the configuration of out-of-plane
component of the local spin at $t=1000t_{0}$ for the three cases.
}\label{f7}
\end{figure}

In order to study the movement of Fe DW driven by current, here we
consider a one dimension spin chain model, where the local spin of
DW is assumed to be uniform in the plane perpendicular to the
transport direction. According to the LLG equation and the
nonadiabatic STT modeled above, the dynamical equation for the unit
vector of local spin in layer $i$ is
\begin{align}
\frac{\partial\mathbf{n}_{i}}{\partial t}=-\gamma\mathbf{n}_{i}\times\mathbf{H}%
_{\mathrm{eff},i}-\alpha \mathbf{n}_{i}\times
\frac{\partial\mathbf{n}_{i}}{\partial t}%
-b_{J}\frac{\partial\mathbf{n}_{i}}{\partial y}\nonumber\\
-\beta
b_{J}\mathbf{n}_{i}\times\frac{\partial\mathbf{n}_{i}}{\partial
y}+g_{J}\mathbf{n}_{i}\times\frac{\partial^{3}\mathbf{n}_{i}}{\partial
y^{3}} \label{LLG}
\end{align}
where the extra $\beta$ term $-\beta
b_{J}\mathbf{n}_{i}\times\partial_{y}\mathbf{n}_{i}$ is only for
comparision. Here $i$ is the site index of one dimension spin chain
and the distance between two sites is $d$, which is unit length used
in our dynamical calculations. In the meanwhile, we also use the
dimensionless time where the unit time is $t_{0}=(4\pi
M_{s}\gamma)^{-1}$. The Gilbert damping coefficient is chosen as
$\alpha=0.02$ and $\gamma$ is the gyromagnetic ratio.
$\mathbf{H}_{\mathrm{eff},i}$ is effective field for the local spin
on site $i$, which is written as\cite{ohe-prl06}
\begin{align}
\mathbf{H}_{\mathrm{eff},i} =H_{K}n_{z,i}\mathbf{e}_{z}+4\pi M_{s}H_{ex}\left( \mathbf{n}%
_{i-1}+\mathbf{n}_{i+1}\right)-4\pi M_{s}n_{x,i}\mathbf{e}_{x}
\label{heff}
\end{align}
where $H_{ex}$ is dimensionless exchange constants and $H_{K}$ is
dimensionless anisotropy field. For the equilibrium DW without
current, the DW width is determined by
$\lambda_{\mathrm{DW}}=\sqrt{H_{ex}/H_{K}}$[ML]. Here we take
$H_{ex}=3.0$ and $H_{K}=0.03$ so that the width of Fe DW in our
simulation is 10[ML]. In addition, in our calculations the current
density is $J=10^{11}[\mathrm{A}/\mathrm{m}^{2}]$, so the
coefficients in STT terms based on our first-principles calculated
results are $b_{J}=0.0154$ and $g_{J}=0.093$, which are measured in
unit time $t_{0}$ and unit length $d$. Further, in order to
demonstrate the difference of DW movement between the case of
$\beta$ term and the third-order STT, we also take $\beta=-0.12$,
which is chosen to make the maximum of $\beta$ term be equal to that
of the third-order STT. Hence, given the initial configuration of
DW, the above LLG equations for each unit vector of local spin can
be solved numerically by Runge-Kutta method.

The position of Fe DW center as function of time with adiabatic and
two kinds of nonadiabatic STTs are shown in Fig. \ref{f7}. As the
prediction by the previous studies,\cite{Li-prb04,Li-prl04} our
calculated result for the case of only adiabatic STT ($\beta=0,
g_{J}=0$) shows that the DW center start to move at first and then
will stop inevitably at last. This phenomenon is attributed to the
existence of out-of-plane component of local
spin,\cite{stiles-prb07,tatara-prl04} i.e., the local spin will be
tilted in $x$ direction after the current have been injected into
the DW. As shown in the inset of Fig. \ref{f7}, one can observe that
the out-of-plane component of DW local spin moments in this case
will raise up when the DW center stopped. Owing to the shape of DW
is film-like, the out-of-plane component of local spin moments can
produce the demagnetization field
$\mathbf{H}_{\mathrm{demag},i}=-4\pi M_{s}n_{x,i}\mathbf{e}_{x}$,
which can produce a torque resisting the adiabatic in-plane STT.
Therefore, once the demagnetization field increase to be large
enough, the velocity of DW center will decrease to zero at last.

The DW movements driven by the nonadiabatic STT of the third-
($\beta=0, g_{J}=0.093$) and first-order ($\beta=-0.12, g_{J}=0$)
term are also shown in Fig. \ref{f7}. It can be seen that in both
cases the DW will no longer stop and the velocity is nearly constant
as the DW center move away from its initial position. The DW
velocity induced by the third-order STT is about half of wall speed
driven by the $\beta$ term. From Fig. \ref{f7} our obtained wall
speed caused by $\beta$ term is about 5.0[m/s], which is exactly
equal to the value of $v=b_{J}\beta/\alpha$. Note that the DW center
move backward in contrast to the case of adiabatic STT. In the inset
of Fig. \ref{f7}, we also show the configuration of the out-of-plane
component of local spin at $t=1000t_{0}$ for these two cases.
Compared with the case of only adiabatic STT, the out-of-plane
component of local spin moments with third- and first-order STT are
much greater. So the precession around the relative larger
demagnetization field will not only resist the adiabatic in-plane
STT but also can push the DW moving backward that is in the opposite
direction for the case of only adiabatic STT. Moreover, as shown in
Fig. \ref{f5} the area under the nonadiabatic STT distribution curve
for the third-order term is about half as much as that for the
$\beta$ term. Therefore, one can see that the out-of-plane component
of local spin at DW center for the third-order STT is also half of
the value for the case of $\beta$ term. As consequence, it will lead
to about 50\% decrease in the values of both corresponding
demagnetization field and velocity for third-order STT compared to
the case of $\beta$ term.

\section{Summary}

In summary, we have calculated the current-induced adiabatic and
nonadiabatic STTs of Fe, Co and Ni DWs by first-principles
noncollinear scattering wave function matching method in the frame
of TB-LMTO with ASA approximation. We found that in Co and Ni DWs
the spatial distribution of the nonadiabatic STT are in the form of
typical nonlocal oscillating torques. However, in contrast to the
case of Co and Ni DWs, the out-of-plane component of nonadiabatic
STT in Fe DW has localized form. The calculated results also show
that the degree of nonadiabaticity in STT of Fe DW is much larger
than that of Co and Ni DWs. Further, we found that the distribution
of localized nonadiabatic STT in Fe DW can be well fitted by the
term of $g_{J}\mathbf{n}\times\partial^{3}_{y}\mathbf{S}$, which is
in the third-order spatial derivative of the local spin instead of
the well-known first-order nonadiabatic $\beta$ term. The
coefficient $g_{J}$ is also obtained for different width
$\lambda_{\mathrm{DW}}$ in Fe DW. Finally, the dynamics of local
spin in Fe DW driven by the third-order nonadiabatic STT is
calculated using LLG equation. The results show that with the same
amplitude of nonadiabatic STT, the velocity of DW center driven by
the third-order STT is about half of wall speed caused by the
first-order $\beta$ term.

\section*{Acknowledgments}

The authors acknowledge Prof. Ke Xia for suggesting the problem and
Dr. Shuai Wang for useful discussion about the calculations. The
authors are also grateful to Dr. Yuan Xu, Dr. Yong Wang and Dr. Rui
Wang for technical assistance. We are grateful to: Ilja Turek for
his TB-LMTO-SGF layer code; Anton Starikov for the TB-MTO code based
upon sparse matrix techniques. The authors also acknowledge the
financial support from the National Natural Science Foundation of
China (Grant No:11104247), China Postdoctoral Science Foundation
(Grant No:2012M520666), and the Provincial Natural Science
Foundation of Zhejiang (Grant No: Y201121807 and Y13A040032).

\end{document}